\documentclass[aps,prb,reprint,superscriptaddress,floatfix]{revtex4-1}

\usepackage{graphicx}
\usepackage{hyperref}
\usepackage{textcomp} 

\usepackage[english]{babel}
\usepackage{blindtext}
\usepackage{color}
\usepackage{soul} 
\usepackage{placeins}
\usepackage{siunitx}
\usepackage{changes}
\usepackage{makecell}

\begin{document}

\title{A Mg-pair isoelectronic bound exciton identified by its isotopic fingerprint in $^{28}$Si}
\author{R. J. S. Abraham}
\author{A. DeAbreu}
\author{K. J. Morse}
\affiliation{Department of Physics, Simon Fraser University, Burnaby, British Columbia, Canada V5A 1S6}
\author{V. B. Shuman}
\affiliation{Ioffe Institute, Russian Academy of Sciences, 194021 St. Petersburg, Russia}
\author{L. МM. Portsel}
\affiliation{Ioffe Institute, Russian Academy of Sciences, 194021 St. Petersburg, Russia}
\author{АA. N. Lodygin}
\affiliation{Ioffe Institute, Russian Academy of Sciences, 194021 St. Petersburg, Russia}
\author{Yu. A. Astrov}
\affiliation{Ioffe Institute, Russian Academy of Sciences, 194021 St. Petersburg, Russia}
\author{N. V. Abrosimov}
\affiliation{Leibniz Institute for Crystal Growth, 12489 Berlin, Germany}
\author{S. G. Pavlov} 
\affiliation{Institute of Optical Sensor Systems, German Aerospace Center (DLR), 12489 Berlin, Germany}
\author{H.-W. H$\ddot{\text{u}}$bers}
\affiliation{Institute of Optical Sensor Systems, German Aerospace Center (DLR), 12489 Berlin, Germany}
\affiliation{Humboldt Universit$\ddot{\text{a}}$t zu Berlin, Department of Physics, 12489 Berlin, Germany}
\author{S. Simmons}
\affiliation{Department of Physics, Simon Fraser University, Burnaby, British Columbia, Canada V5A 1S6}
\author{M. L. W. Thewalt}
\email[Corresponding author: ]{thewalt@sfu.ca}
\affiliation{Department of Physics, Simon Fraser University, Burnaby, British Columbia, Canada V5A 1S6}

\date{\today}

\begin{abstract}
We use the greatly improved optical linewidths provided by highly enriched $^{28}$Si to study a photoluminescence line near $\SI{1017}{meV}$ previously observed in the luminescence spectrum of natural Si diffused with Mg, and suggested to result from the recombination of an isoelectronic bound exciton localized at a Mg-pair center. In $^{28}$Si this no-phonon line is found to be comprised of five components whose relative intensities closely match the relative abundances of Mg-pairs formed by random combinations of the three stable isotopes of Mg, thus confirming the Mg-pair hypothesis. We further present the results of temperature dependence studies of this center that reveal unusual and as yet unexplained behaviour.
\end{abstract}
\maketitle

\section{Introduction}
Deep double donors in silicon have recently been proposed as the basis for a scalable spin-qubit/photonic-cavity technology \cite{Morse2017}. Our recent reinvestigation \cite{Abraham2018} of the Lyman absorption transitions of the deep double donor interstitial magnesium (Mg$_i$) in silicon clarified a number of issues raised in earlier studies \cite{Ho1972,Ho1993,Ho1998,Ho2003,2Ho2003,Ho2006} of mid-infrared absorption in Si diffused with Mg. The dominant interstitial location of Mg, acting as a double donor, is unusual, with other Group-II elements such as zinc and beryllium forming primarily substitutional double acceptors in Si \cite{Carlson1957, Robertson1968, Crouch1972}. There is also evidence that magnesium may occupy a substitutional site in Si and act as a deep double acceptor (Mg$_s$), with deep-level transient spectroscopy on Mg-diffused samples showing hole-emission attributed to Mg$_s$ \cite{Baber1988}. \par

The dual character of magnesium as a substitutional double acceptor and an interstitial double donor implies the potential for formation of an isoelectronic bound exciton (IBE) center comprised of an Mg$_i$-Mg$_s$ pair, similar to the well-known $\SI{1077}{meV}$ Be-pair IBE \cite{Henry1981}. IBE centers are able to localize excitons much more tightly than shallow donors or acceptors, and thanks to their lack of nonradiative Auger decay, can provide very bright luminescence even when present at low concentrations. The availability of highly isotopically enriched $^{28}$Si, together with its greatly reduced optical linewidths, due to the near-elimination of inhomogeneous broadening, \cite{Cardona2005} makes possible a new method of determining the chemical constituents of an IBE binding center. This method, known as isotopic fingerprinting \cite{Steger2008, Steger2011}, is an extension of the well-known method of no-phonon (NP) line isotope shifts \cite{Heine1975}. The narrow linewidths in $^{28}$Si result in distinct spectral components for all possible combinations of the stable isotopes comprising the IBE binding center, instead of just the shift of a relatively broad line which is seen in natural Si. \par

The photoluminesence (PL) results presented in this work follow the results of Steinman and Grimmeiss \cite{Steinman1998}, who proposed that an Mg$_i$-Mg$_s$ IBE pair center might be responsible for an NP line observed in PL at $\SI{1017}{meV}$. Here we confirm this hypothesis through high resolution PL spectroscopy of $^{28}$Si diffused with natural Mg. The $\SI{1017}{meV}$ NP line is revealed to comprise five distinct peaks with relative intensities matching those anticipated for a center containing two Mg atoms given the abundances \cite{CRC2016} of the three stable naturally occurring Mg isotopes: 78.99(4)$\%$ $^{24}$Mg, 10.00(1)$\%$ $^{25}$Mg and 11.01(3)$\%$ $^{26}$Mg. Temperature dependence measurements of the PL of this IBE center revealed unusual behaviour of the main NP line and associated local vibrational mode (LVM) and phonon replicas. \par

\section{Materials and methods}
The sample studied here was examined in our previous \cite{Abraham2018} study of absorption transitions in Mg-doped Si, where it was referred to as $^{28}$Si low boron (LB). Shuman et al.\cite{1Shuman2017, 2Shuman2017} have detailed the methods used for the diffusion of Mg into Si. The starting material was enriched to 99.995$\%$ $^{28}$Si, and has been thoroughly described elsewhere \cite{Devyatykh2008}. All PL measurements were performed using a Bruker IFS 125HR Fourier transform infrared (FTIR) spectrometer with a CaF$_2$ beam-splitter. Samples were mounted loosely so as to avoid strain and cooled in a liquid helium cryostat. PL was generated using $\SI{1030}{nm}$ and $\SI{1047}{nm}$ excitation sources for high resolution scans of the IBE and temperature-dependent PL measurements, respectively. Scattered excitation light was removed by sharp-cut long-pass filters. For PL measurements at temperatures above $\SI{4.2}{K}$, the sample was lightly held against a temperature-controlled Cu plate cooled in flowing He gas. The PL was detected with a liquid nitrogen cooled Ge PIN diode detector. For the high resolution isotopic fingerprint PL scans an apodized instrumental resolution of $\SI{3.1}{\micro eV}$ full width at half maximum (FWHM) was used, while a lower resolution of $\SI{124}{\micro eV}$ FWHM was used for the temperature dependence study, since for these studies lower excitation levels were necessary to minimize sample heating when the sample was not immersed in liquid He.

\section{Results}
\FloatBarrier
\subsection{Isoelectronic bound exciton fine structure}
\begin{figure}[htp]
\includegraphics[width=0.41\textwidth]{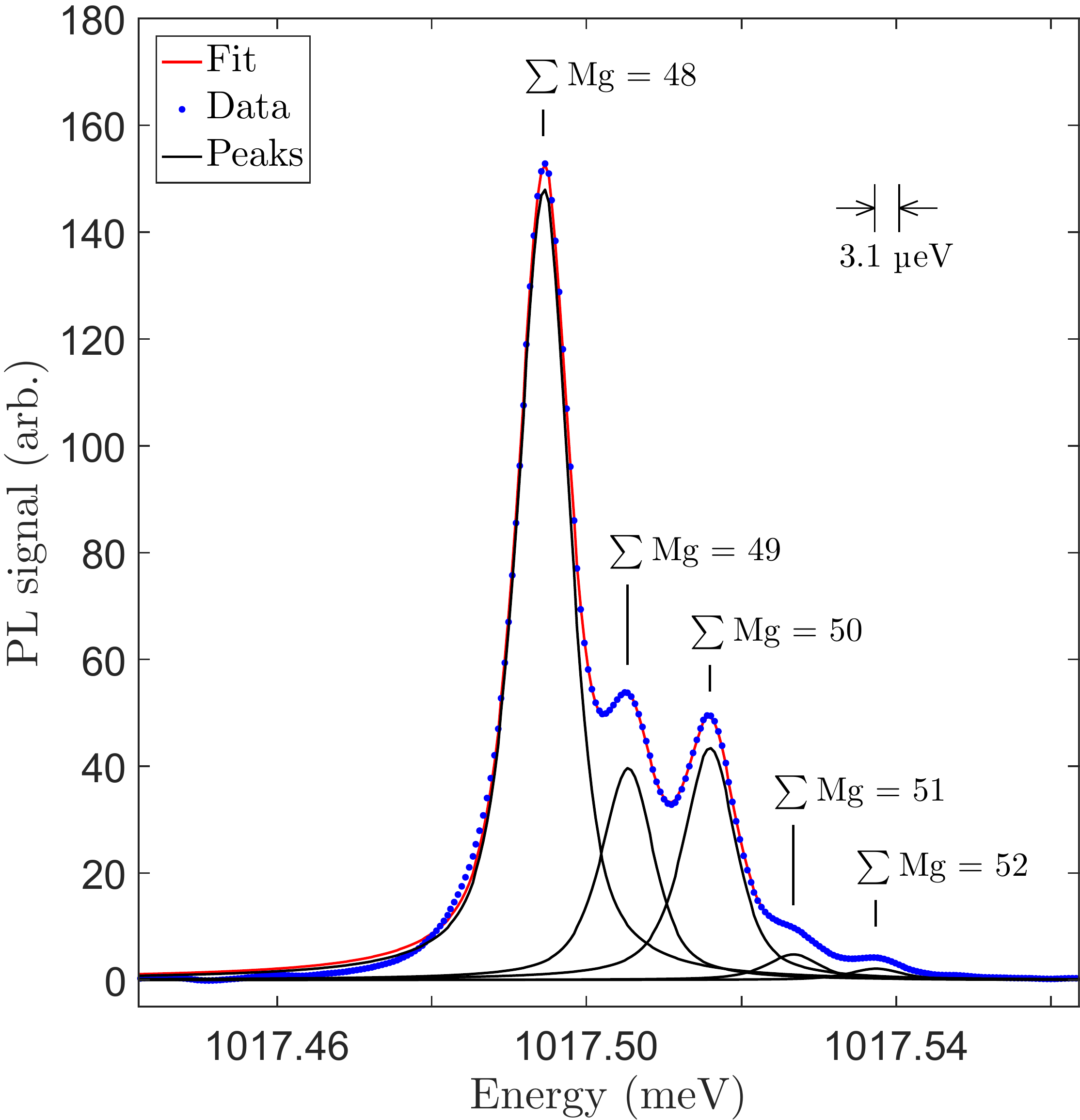}
\caption{Isotopic fingerprint of the magnesium isoelectronic pair center. The mass of the Mg-pair center in amu corresponding to each possible combination of naturally occurring isotopes is labelled. Spectra were collected at a resolution of $\SI{3.1}{\micro eV}$ with $\SI{1031}{nm}$ laser excitation.}
\label{PLfingerprint}
\end{figure}

In the PL spectrum of the Mg-doped $^{28}$Si LB sample we observed a relatively strong NP feature at $\SI{1017}{meV}$, which we label Mg$_{NP}$. This had been observed previously by Steinman and Grimmeiss \cite{Steinman1998} who suggested it might arise from a substitutional-interstitial Mg pair, in analogy with the well-known Be-pair IBE luminescence center \cite{Henry1981}. Thanks to the near-elimination of inhomogeneous broadening made possible with $^{28}$Si, the NP line is seen to resolve into five components, as shown in Fig.~\ref{PLfingerprint}. The fit to the fine structure seen in Fig.~\ref{PLfingerprint} consists of five mixed Gaussian-Lorentzian peaks with an asymmetry parameter \cite{Aaron2008} that is kept constant for all peaks. This slight asymmetric broadening to low energy may be a consequence of Stark broadening due to random electric fields present even when using relatively lightly compensated starting material, due to Mg incorporating as both a substitutional (acceptor) and interstitial (donor) impurity. \par 

The relative intensities of the five observed components closely match the expected relative abundances of Mg-Mg pairs with total masses ranging from 48 to 52 amu, as shown   in Tab.~\ref{PLtable}. The fact that the observed spectrum can be explained by grouping together isotope combinations having the same total mass indicates that the Mg atoms are indistinguishable from each other in terms of the NP isotope shifts, which is quite a common occurrence for other IBE centers in $^{28}$Si whose isotopic fingerprints have been studied \cite{Steger2011}. This result provides very strong confirmation for the proposal that the $\SI{1017}{meV}$ NP line observed in Mg-doped Si arises from a center containing a pair of Mg atoms. \par

\begin{table}[htp]
\begin{ruledtabular}
\begin{tabular}{l c c c}
& \multicolumn{2}{c}{Integrated intensity} \\
\cline{2-3}
$\sum$Mg (amu) & Observed & Predicted & Peak energy (meV)\\
\hline
48 & 0.62(3) & 0.6239(6) & 1017.494\\
49 & 0.17(1) & 0.1580(8) & 1017.505\\
50 & 0.18(1) & 0.184(1) & 1017.516\\
51 & 0.020(3) & 0.0220(6) & 1017.527\\
52 & 0.009(2) & 0.0121(4) & 1017.537\\
\end{tabular}
\end{ruledtabular}
\caption{Observed and predicted relative integrated intensities for peaks corresponding to Mg-Mg pairs. Total amu ranges from 48 to 52 depending on the combination of isotopes with $^{24}$Mg-$^{24}$Mg comprising the largest fraction and $^{26}$Mg-$^{26}$Mg the smallest. Absolute values of peak position are deemed accurate to two decimal places. Relative positions of the peaks to one another are considered valid to three decimal places.}
\label{PLtable}
\end{table}

\FloatBarrier
\subsection{Temperature dependence}
\begin{figure}[htp]
\includegraphics[width=0.45\textwidth]{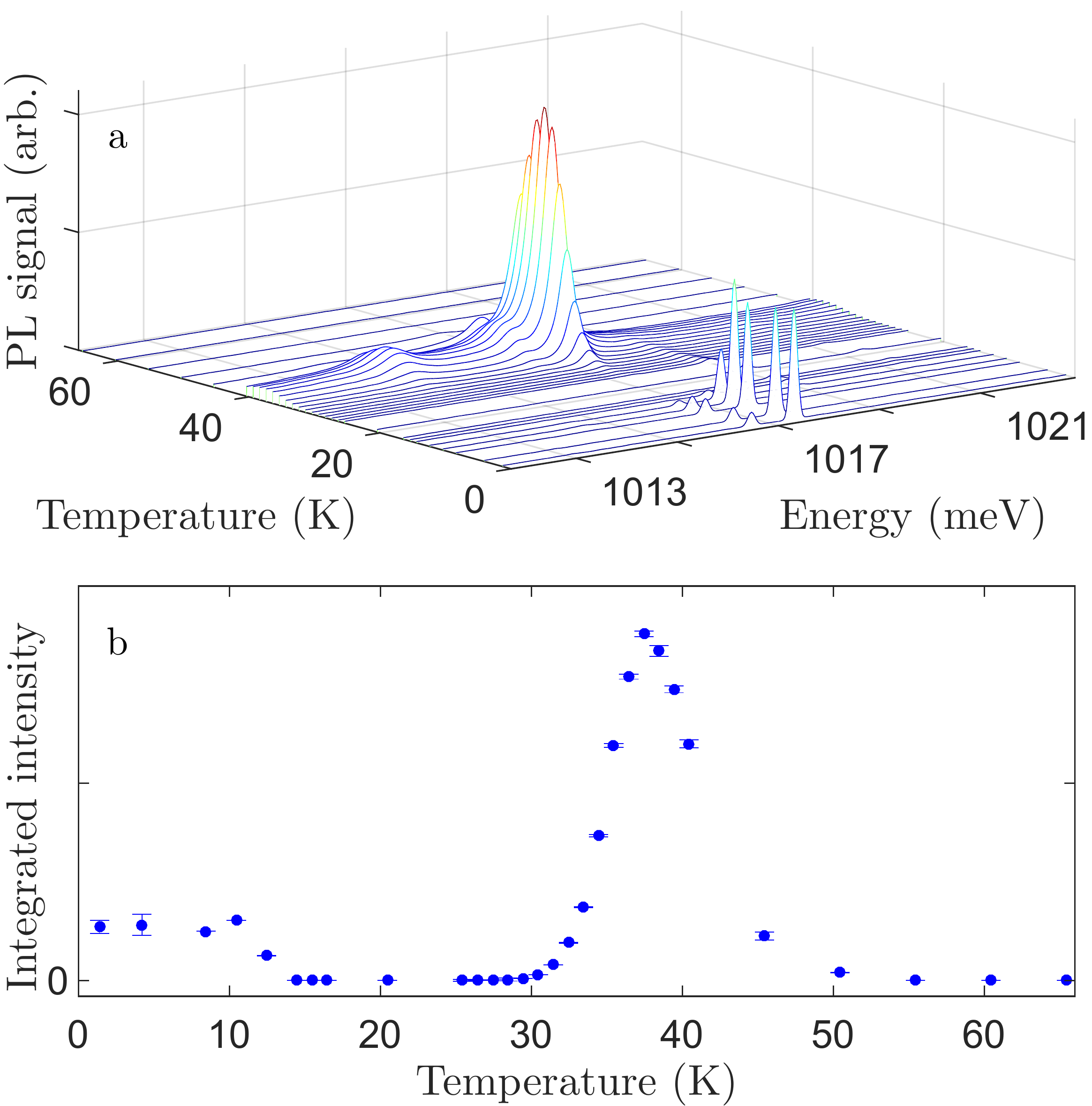}
\caption{\textbf{(a)} Waterfall plot showing the temperature dependence of the Mg-Mg IBE line and replicas. Up to $\sim\!\SI{10}{K}$ the IBE features are relatively unchanged, but then decrease rapidly with increasing temperature until they disappear completely above $\sim\!\SI{17}{K}$. A resurgence of the IBE PL with  increased linewidth and intensity along with a new LVM phonon replica become visible above $\sim\!\SI{30}{K}$. Spectra were collected at a resolution of $\SI{124}{\micro eV}$ with $\SI{1047}{nm}$ laser excitation. \textbf{(b)} Integrated intensity of the Mg$_{NP}$ line as a function of temperature.}
\label{PLwaterfall}
\end{figure}

\begin{figure}[htp]
\includegraphics[width=0.41\textwidth]{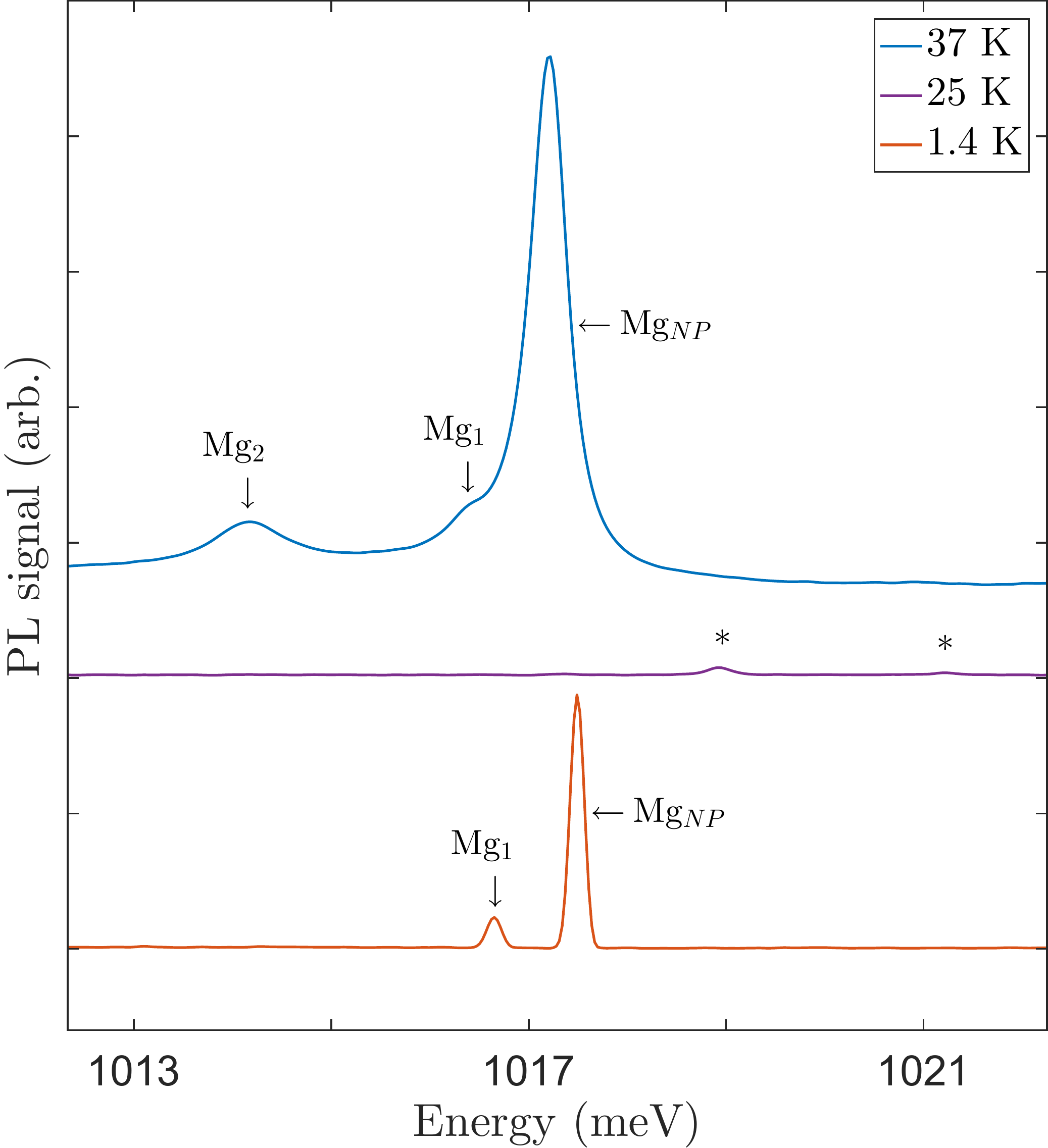}
\caption{Example spectra showing the Mg$_{NP}$ line and associated features Mg$_1$ and Mg$_2$ in the high, intermediate, and low temperature regimes. Two weak features labelled $*$ in the $\SI{25}{K}$ spectrum are of unknown origin. Spectra were collected at a resolution of $\SI{124}{\micro eV}$ with $\SI{1047}{nm}$ laser excitation.}
\label{PLselectedtemps}
\end{figure}

Temperature dependence studies of the isoelectronic Mg pair center above $\SI{4.2}{K}$ show unusual behaviour as summarized in Fig.~\ref{PLwaterfall}. At liquid He temperatures the Mg$_{NP}$ line is strong and very sharp, allowing the isotopic fingerprint to be resolved at both $\SI{1.4}{K}$ and $\SI{4.2}{K}$, and a sharp line which we label Mg$_1$ can be observed $\SI{0.84}{meV}$ below the NP line. As $\SI{20}{K}$ is approached both the NP line and the Mg$_1$ line vanish, only to reappear above $\SI{30}{K}$ with substantially increased linewidth and intensity, together with a new feature downshifted from the NP line by $\SI{3.0}{meV}$, which we label Mg$_2$. There is no sign of the Mg$_2$ line in the low temperature spectra. Fig.~\ref{PLselectedtemps} compares the IBE spectrum at three temperatures best illustrating this unusual behaviour. Two weak features labelled $*$ in the $\SI{25}{K}$ intermediate temperature spectrum shown in Fig.~\ref{PLselectedtemps} are of uncertain origin. The PL spectrum of this sample contains a number of weak, unidentified sharp features which may result from unintentional impurities introduced during the Mg-diffusion, and/or the complexes of such impurities with Mg. The rapid decrease in IBE PL intensity above $\sim\!\SI{37}{K}$, seen most clearly in Fig.~\ref{PLwaterfall} \textbf{b}, is very typical of IBE centers, and results from thermal dissociation of whichever electronic particle, the electron or the hole, that is weakly bound by Coulomb attraction to the oppositely charged more tightly bound particle. \par 

\begin{figure}[htp]
\includegraphics[width=0.41\textwidth]{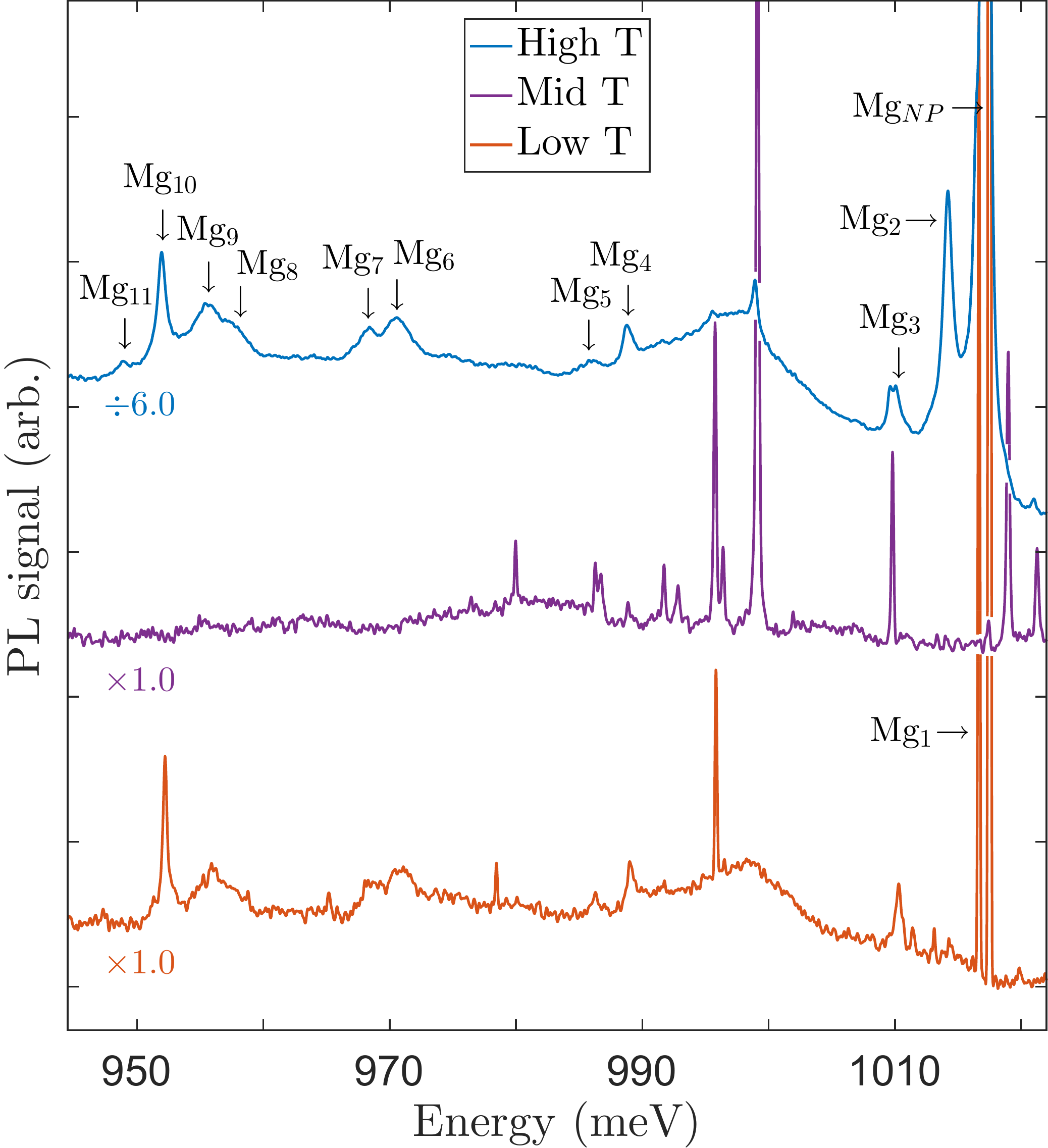}
\caption{PL spectra of the main Mg$_{NP}$ line and a range of features below it in energy in three temperature regimes. For improved SNR we have averaged spectra at 35, 36 and $\SI{37}{K}$ to show the high temperature regime. Similarly, spectra at 16, 20, and $\SI{25}{K}$ are averaged to show the middle regime, and spectra at 1.4 and $\SI{4.2}{K}$ are averaged to show the low temperature regime. We note a number of features, labelled Mg$_1$-Mg$_{11}$, that display a similar temperature dependence to the main NP line that are likely related to the Mg-pair IBE. Many relatively narrow PL features of unknown origin, but with a different temperature dependence, most notable in the mid T regime, are also observed but not labelled. The high temperature spectrum has been scaled down by a factor of 6.0 to make the areas of the Mg$_{NP}$ line in the high and low temperature spectra approximately equal for ease of comparison of the replicas. For both the high and low temperature spectra, the peak of the main Mg$_{NP}$ line is truncated at the top edge of the plot, and the peak of the Mg$_1$ line in the low temperature spectrum lies at the top edge of the plot. Spectra were collected at a resolution of $\SI{124}{\micro eV}$ with $\SI{1047}{nm}$ laser excitation. The temperature dependence of the Mg$_1$ and Mg$_2$ features relative to the Mg$_{NP}$ line is better seen in Fig.~\ref{PLselectedtemps}.}
\label{IBE_lowEnergy}
\end{figure}

Specific phonon and LVM replicas are very characteristic of different IBE centers, and these have not been reported previously for the Mg-pair IBE. This identification is complicated by the fact that the Mg-pair IBE luminescence in our sample does not dominate the PL spectrum as do many other IBE centers in optimized samples, and the PL spectrum of this Mg-diffused sample contains many unidentified weak but sharp lines. The replicas of the Mg$_{NP}$ line are shown over a wider energy region in Fig.~\ref{IBE_lowEnergy}. Many unlabelled features, particularly in the intermediate temperature spectrum where the Mg-pair IBE features vanish, likely result from other impurities introduced during the Mg diffusion or from complexes containing more than two Mg atoms. \par 

\begin{table}[htp]
\begin{ruledtabular}
\begin{tabular}{l c}
Label & Shift from Mg$_{NP}$ (meV) \\
\hline
Mg$_1$ & 0.84\\
Mg$_2$ & 3.02\\
Mg$_3$ & 7.2\\
Mg$_4$ & 28.4\\
Mg$_5$ & 31.2\\
Mg$_6$ & 46.6\\
Mg$_7$ & 49.0\\
Mg$_8$ & 59.3\\
Mg$_{9}$ & 61.6\\
Mg$_{10}$ & 65.28\\
Mg$_{11}$ & 68.3\\
\end{tabular}
\end{ruledtabular}
\caption{Energy shifts below the Mg$_{NP}$ line of luminescence components which are believed to be related to the Mg-pair IBE.  The Mg$_{11}$ replica, which like Mg$_2$ is only observed in high temperature spectra, likely results from the emission of the modes responsible for the Mg$_2$ replica plus the mode responsible for the Mg$_{10}$ replica.}
\label{lowenergy_PLtable}
\end{table}

The unusual temperature dependence of the Mg-pair IBE PL can be used to associate features which are replicas of the Mg$_{NP}$ line, as already shown in Fig.~\ref{PLselectedtemps} for the Mg$_1$ and Mg$_2$ lines, and reject unknown features which behave differently. In Table \ref{lowenergy_PLtable} we summarize the energy shifts of these lower-energy features which we believe are associated with the Mg-pair IBE. With a shift of only $\SI{0.84}{meV}$, the Mg$_1$ feature is unlikely to result from a vibronic replica, but may instead originate in another electronic state of the Mg-pair IBE. The Mg$_2$ feature, which appears only in the high temperature regime, has a shift which is more compatible with LVM replicas seen for other IBE in silicon \cite{Steger2008}. The strong Mg$_{10}$ replica has a shift of $\SI{65.28}{meV}$, very close to the $\SI{65.03}{meV}$ energy of the zone-center optical phonon energy measured using Raman scattering in these $^{28}$Si samples. \par

The identification of Mg$_3$ as a replica of the Mg-pair IBE is tentative, as an unknown line, seen in the intermediate temperature spectrum of Fig.~\ref{IBE_lowEnergy}, is superimposed on it. The Mg$_{11}$ feature, which is observed only in the high temperature spectra, may be a combination of the emission of the modes responsible for Mg$_2$ and Mg$_{10}$. \par

\FloatBarrier
\section{Conclusion and Outlook}
We have studied the PL spectrum of highly enriched $^{28}$Si diffused with Mg to reexamine a luminescence line near $\SI{1017}{meV}$ previously observed in the PL spectrum of natural Si doped with Mg by Steinman and Grimmeiss \cite{Steinman1998} and proposed by them to be an IBE localized on a Mg-pair center. The isotopic fingerprint of this center made possible by the near-elimination of inhomogeneous broadening in $^{28}$Si provides strong confirmation that the binding center of this IBE contains two Mg atoms. A number of lower energy replicas of the main Mg$_{NP}$ line were identified. \par 

The behaviour of the PL of this IBE center with changing sample temperature is very unusual, and to the best of our knowledge is unique in Si. From pumped He temperature to $\sim\!\SI{10}{K}$ there is little change in the spectrum, but from $\sim\!\SI{10}{K}$ to $\sim\!\SI{17}{K}$ the intensity of all components decreases and vanishes from $\sim\!\SI{20}{K}$ to $\sim\!\SI{30}{K}$. Above $\sim\!\SI{30}{K}$ they reappear with increased intensity and NP linewidth, and a new replica downshifted by $\SI{3.02}{meV}$ appears which cannot be observed in the low temperature spectrum. Above $\sim\!\SI{37}{K}$ all components decrease in intensity, and disappear above $\sim\!\SI{50}{K}$. This last behaviour is familiar for all IBE, and results from the thermal dissociation of the weakly-bound electronic particle of the IBE. \par 

The disappearance of all IBE PL between $\sim\!\SI{20}{K}$ and $\sim\!\SI{30}{K}$, and its reappearance above $\sim\!\SI{30}{K}$ with increased intensity and NP linewidth, and with a new LVM replica, at present has no explanation. Further studies of the Mg-pair IBE would benefit from a careful optimization of the diffusion and annealing conditions which could maximize its intensity. Maximizing the total Mg concentration is known to not be optimal, as such samples have extremely weak PL, likely due to nonradiative recombination at Mg-precipitates as suggested by Steinman and Grimmeiss \cite{Steinman1998}. \par

\FloatBarrier
\section{Acknowledgements}
This work was supported by the Natural Sciences and Engineering Research Council of Canada (NSERC), the Canada Foundation for Innovation (CFI), and the British Columbia Knowledge Development Fund (BCKDF). This work has been partly supported by the Russian Foundation for Basic Research (RFBF Project No. 18- 502-12077-DFG) and of the Deutsche Forschungsgemeinschaft (DFG No. 389056032). The $^{28}$Si samples used in this study were prepared from Avo28 crystal produced by the International Avogadro Coordination (IAC) Project (2004-2011) in cooperation among the BIPM, the INRIM (Italy), the IRMM (EU), the NMIA (Australia), the NMIJ (Japan), the NPL (UK), and the PTB (Germany). \par

\end{document}